\documentstyle[pra,aps,epsf]{revtex}
\def\fsz{\footnotesize}
\def\comment#1{}

\begin{document}
\title{Convergence behavior of variational perturbation expansion---\\
A method for locating
Bender-Wu singularities}
\author{
H. Kleinert$^1$ and
W. Janke$^{2}$
%
}
\address{$^1$ Institut f\"{u}r Theoretische Physik,
        Freie Universit\"{a}t Berlin,
        Arnimallee 14, 14195 Berlin, Germany\\
        $^2$Institut f\"{u}r Physik,
        Johannes Gutenberg-Universit\"at Mainz,
        Staudinger Weg 7, 55099 Mainz, Germany
       }
\date{\today}
\maketitle
\begin{abstract}
Variational perturbation expansions have recently
been used to calculate directly the strong-coupling expansion coefficients
of the anharmonic oscillator. The convergence is exponentially
fast with superimposed oscillations, as recently
observed empirically
by the authors.
In this note, the observed
behavior is explained and used to determine accurately the
magnitude and phase of the leading
Bender-Wu singularity which is responsible
for the finite convergence
radius in the complex coupling constant plane.
\end{abstract}
\begin{figure}[tbhp]
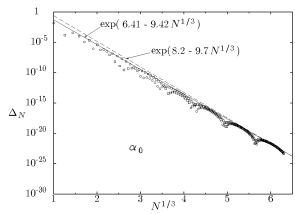

\input  slope1
\caption[Exponentially fast convergence of the $N$th approximants for $ \alpha
_0$
to the exact value]{Exponentially fast convergence
of the $N$th approximants for $ \alpha _0$
to the exact values. The plotted quantity is
$ \Delta _N\equiv |( \alpha _0)_N)- \alpha _0|$.
}
\label{slope1}\end{figure}
\begin{figure}[tbhp]
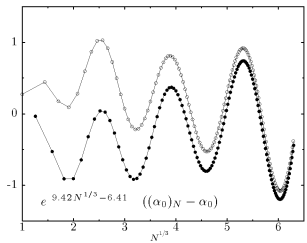

\input  janke12
\caption[Oscillatory behavior around the asymptotic approach
of $ \alpha _0$ to its exact value
as a function of the order $N$ of the approximant]{
Oscillatory behavior around the asymptotic approach
of $ \alpha _0$ to its exact value
as a function of the order $N$ of the approximant (open circles
are for odd $N$, filled circles for even $N$).}
\label{janke12}\end{figure}

\noindent
Variational perturbation theory
yields uniformly and exponentially fast converging expansions for
many quantum
mechanical systems
\cite{syko,PI,sz}.
The convergence was proved for the anharmonic integral and
the quantum mechanical anharmonic oscillator
in several papers for a {\em finite}
coupling strength $g$ of the anharmonic term \cite{conv}.
Recently, this theory
has been adapted to
calculating
the coefficients of strong-coupling expansions \cite{jk}. For the
ground-state energy of the anharmonic oscillator, the first 23 coefficients
of the series
\begin{eqnarray}
E^{(0)}=  \left(\frac{g}{4}\right)^{1/3}\left[ \alpha _0
+ \alpha _1\left(\frac{g}{4 \omega ^3}\right)^{-2/3}
+ \alpha _2\left(\frac{g}{4 \omega ^3}\right)^{-4/3}
+\dots\right]
,
\label{5.scexpag}\end{eqnarray}
were calculated with 20 digits precision.
Thereby an interesting phenomenon was observed:
the approach to the asymptotic values of $ \alpha _n$.
is exponentially fast
with periodic modulations, as shown
in Figs. \ref{slope1} and \ref{janke12} calling for a revision of
the above-quoted
proofs
which do not apply to the strong-coupling limit $g \rightarrow \infty$.
The purpose of this note is
to explain the observed behavior theoretically
and to show that it leads to a rather precise estimate
of the largest Bender-Wu singularities in the complex-$g$-plane which
determine the convergence radius of the strong-coupling expansion.

The potential of the anharmonic oscillator is
\begin{equation}
V(x) = \frac{\omega^2}{2} x^2 + \frac{g}{4} x^4 \hspace{1cm} ( \omega^2,g>0).
\end{equation}
The Rayleigh-Schr\"{o}dinger perturbation theory
yields a power-series expansion
\begin{equation}
   E^{(0)}(g) = \omega \sum_{l=0}^{\infty} E^{(0)}_{l} \left(
   \frac{g/4}{\omega^3} \right)^l,
\label{eq:1}
\end{equation}
where \(E^{(0)}_l\) are rational numbers
   1/2, 3/4, -21/8, 333/16,
   -30885/128, \dots ~,
obtained from the recursion relations of Bender and Wu \cite{bewu}.

As is well known,
the series (\ref{eq:1}) cannot be used for an evaluation of the energy
since it
has a zero
radius of convergence due to the factorial growth of the coefficients
\( E^{(0)}_{l} \).
The recently developed variational perturbation theory \cite{PI}
converts the divergent series (\ref{eq:1}) into an
exponentially fast convergent one.
The procedure goes as follows (see Section 5.13 of Refs. \cite{PI}).
First, the harmonic term of the potential is
split into a new harmonic term with  a trial frequency $ \Omega $,
 and a remainder:
\begin{equation}
   \frac{\omega^2}{2} x^2 = \frac{\Omega^2}{2} x^2 +
   \left(\frac{\omega^2}{2}-\frac{\Omega^2}{2}\right)x^2.
\end{equation}
After rewriting
\begin{equation}
   V(x) = \frac{\Omega^2}{2} x^2 + \frac{g}{4} (-2 \sigma  x^2/ \Omega
+ x^4),
\end{equation}
we perform a perturbation expansion in powers of $g$ at a fixed
$ \sigma = \Omega ( \Omega ^2- \omega ^2)/g $,
\begin{equation}
   E_{N}(g, \sigma ) = \Omega \sum_{l=0}^{N} \varepsilon^{(0)}_{l}( \sigma )
\left( \frac{\hat g}{4} \right)^l,
\label{reexpandeden}
\end{equation}
where $\hat g\equiv g/ \Omega ^3$ is a dimensionless reduced
coupling constant.

The calculation of the new series up to a specific order $N$
requires only little work, being
easily obtained from the ordinary
perturbation series (\ref{eq:1}) by replacing
$\omega$ by \(\sqrt{\Omega^2 -g \sigma / \Omega }\), and by
reexpanding (\ref{eq:1}) in powers of $g$ up to the $N$th order.
This yields the reexpansion coefficients
\begin{equation}
   \varepsilon^{(0)}_{l}( \sigma ) = \sum_{j=0}^{l}  E^{(0)}_{j}
      \left( \begin{array}{c}
              (1 - 3 j)/2 \\ l-j
             \end{array}
      \right)
      (-4 \sigma )^{l-j}.  \label{rec}
\end{equation}
The truncated power series
   $W_{N}(g,\Omega) \equiv  E^{(0)}_ {N} \left(
   g, \sigma  \right)$
is certainly independent of $\Omega$ in the limit $N \rightarrow \infty$.
At any finite order, however, it {\em does} depend on $\Omega$ via $ \sigma $,
the
approximation
having its fastest speed of convergence where it depends least on $\Omega$.
If we denote the order-dependent optimal value of $\Omega$ by $\Omega_{N}$,
the quantity $W_{N}(g,\Omega_{N})$ is the new approximation to $E(g)$
\cite{rem}.
Introducing the reduced frequency
$\hat{\omega} = \omega/\Omega,$
the approximation can be written as
\begin{equation}
W_N = \left( g/\hat{g} \right)^{1/3} w_N(\hat{g},\hat{\omega}^2).
\end{equation}

{}From the approximate energies
$W_N$
 it is
easy to derive simple formulas for
the coefficients of the strong-coupling expansion.
We simply
expand the function
$w_N(\hat g,\hat \omega^2)$
in powers of
$\hat{\omega}^2 = (g/\omega^3)^{-2/3} \hat{g}^{2/3}$ and find
\begin{equation}
W_N = (g/4)^{1/3} \left[ \alpha_0
+ \alpha_1 \left(\frac{g/4}{\omega^3}\right)^{-2/3}\!\!\!\!\!\!
+ \alpha_2 \left(\frac{g/4}{\omega^3}\right)^{-4/3}\!\!\!\!\!\! +\dots \right],
\end{equation}
with the coefficients
\begin{equation}
\alpha_n = \frac{1}{n!} w_N^{(n)}(\hat{g},0) \, (\hat{g}/4)^{(2n-1)/3}.
\label{aln}\end{equation}
Here $w_N^{(n)}(\hat{g},0)$ denotes the $n$'th derivative of
$w_N(\hat g,\hat \omega^2)$ with
respect to $\hat{\omega}^2$ at $\hat{\omega}^2=0$,
which can easily be calculated using (\ref{rec}) \cite{jk},
\begin{equation}
\frac{1}{n!} w_N^{(n)}(\hat{g},0) =
\sum_{l=0}^N (-1)^{l+n} \sum_{j=0}^{l-n} E_{j}^{(0)}
      \left( \begin{array}{c}
              (1 - 3 j)/2 \\ l-j
             \end{array}
      \right)
      \left( \begin{array}{c}
              l-j \\ n
             \end{array}
      \right)
      (-\hat{g}/4)^j.
\end{equation}
%
%
The optimal value of $\Omega_N$
has the $N$-dependence
(see Ref. \cite{PI} and the first of Refs. \cite{sz})
\begin{equation}
\Omega_N^3=g c N \left( 1 + 6.85/N^{2/3} \right),
\label{ome}\end{equation}
where the coefficient $c$ is
$c =
0.186\,047\,272\,\dots~.$
With this $ \Omega _N$, we obtain
the exponentially fast approach to the exact
limit as shown in Figs.
\ref{slope1} and
\ref{janke12}. The exponential
falloff is modulated by oscillations.


To explain this behavior we recall that
the
 ground state energy
satisfies
the subtracted
dispersion relation
\begin{equation} \label{17.dr2a}
E^{(0)}(  g)=
\frac{ \omega }{2} +  \frac{ g}{2 \pi i}\int_0^{-\infty}
\frac{d  g'}
{  g'}
\frac{\mbox{disc} \,E^{(0)}(  g')}
{  g'-  g},
\end{equation}
where
$ \mbox{disc}\,E^{(0)}(  g')$
denotes the discontinuity across the left-hand cut
in the complex  $ g$-plane (below minus above).
An expansion of the integrand in powers of $ g$ yields the
perturbation series
(\ref{eq:1}).
The reexpanded series
(\ref{reexpandeden})
 is obtained from
(\ref{eq:1}) by the above replacement of
$ \omega \longrightarrow  \Omega (1- \sigma \hat g)^{1/2} $
and a reexpansion in powers of $g$.
%
%
%

There is a simple way of obtaining
the same reexpansion
from the dispersion relation
(\ref{17.dr2a}).
Introducing the dimensionless coupling
constant $\bar g\equiv   g/ \omega ^3$,
above replacement
amounts to
\begin{equation}
\bar g\longrightarrow \tilde g(\hat g)\equiv \frac{\hat g}{(1- \sigma \hat
g)^{3/2}}.
\label{17.grep}\end{equation}
Since Eq.~(\ref{17.dr2a}) represents an energy,
it can be written as $ \omega $
times a dimensionless function $\bar E^{(0)}(  \bar g)$.
Apart from the replacement  (\ref{17.grep}) in the argument,
it receives an
 overall
factor
$ \Omega / \omega = (1- \sigma \hat g)^{1/2} $.
If we introduce the reduced
energy
$\hat E(\hat g)\equiv {E(g)}/{ \Omega }$,
which depends only on the reduced coupling constant $\hat g$,
the dispersion relation
(\ref{17.dr2a}) for $E^{(0)}(g)$ turns into
\begin{equation}
\hat E^{(0)}(  g)=(1- \sigma \hat g)^{1/2} \left[
\frac{ 1 }{2} +  \frac{\tilde g(\hat g)}{2 \pi i}
\int_0^{-\infty}\frac{ d \bar g'}{\bar g'}
\frac{ \mbox{disc}\, \bar E^{(0)}( \bar g')}
{\bar g'-\tilde g(\hat g)}\right].
 \label{17.drhat}
\end{equation}
The resummed perturbation series is obtained
from this by an expansion in powers
of $\hat g/4$ up to order $N$.

Note
that only the
truncation of the expansion causes a difference between
the two expressions (\ref{17.dr2a})  and  (\ref{17.drhat}),
since $\bar g$ and $\tilde g$ are the same numbers, as can be verified
by inserting $\hat g=g/ \Omega ^3$
and $ \sigma $ into the right-hand side of (\ref{17.grep}).

To find the
reexpansion coefficients
we observe that
the expression (\ref{17.drhat})
satisfies
a dispersion relation in
the complex $\hat g$-plane.
If $C$ denotes the cuts in this plane
and $ \mbox{disc}_CE(\hat g)$
is the discontinuity across these cuts,
the dispersion relation reads
\begin{equation} \label{17.dr2}
\hat E^{(0)}(\hat g)= \frac{1 }{2} +\frac{\hat g}{2 \pi i}
 \int_{C} \frac{d\hat g'}
{\hat g'}
\frac{
\mbox{disc}_C \hat E^{(0)}(\hat g')
}{\hat g'-\hat g}.
\end{equation}
We have changed the argument of the energy
from $\bar g$ to
$\hat g$ since this will be the
relevant variable
in the sequel.

When expanding the denominator in the integrand
in powers of $\hat g/4 $,
the expansion  coefficients $ \varepsilon ^{(0)}_l$
are found to be moment integrals
with respect to the inverse coupling constant $1/\hat g$:
\begin{equation} \label{17.moments}
 \varepsilon^{(0)} _k=\frac{4^k}{2{\pi}i}\int_C
\frac{d\hat g}{{\hat g^{k+1}}}
\mbox{disc}_C\hat E^{(0)}(\hat g).
\end{equation}
In the complex $\hat g$-plane,
the integral (\ref{17.drhat}) has in principle
cuts
along the contours  $C_1,C_{\bar 1},C_2,C_{\bar 2}$, and $C_3$,
 as shown in Fig.~\ref{17.contour}.
The first four cuts are the images of the left-hand cut in the complex
$g$-plane;
the curve $C_3$ is due to the square root of $1- \sigma \hat g$ in
the mapping (\ref{17.grep}) and
the prefactor of (\ref{17.drhat}).

Let $\bar D(\bar g)$ abbreviate the reduced discontinuity
in the original dispersion relation
(\ref{17.dr2a}):
\begin{equation}
\bar D(\bar  g)\equiv {\rm disc} \bar E^{(0)}(\bar g)=
2i{\rm Im\,}\bar E^{(0)}(\bar g-i \eta ),~~~~\bar g\le 0.
\label{17.disgpr}\end{equation}
Then the discontinuities across the various cuts
are
\begin{eqnarray}
\!\!\!\!\!\!\!\!
\!\!\!\!\!\!\!\!\!\!\!\!
\!\!\!\!\!\!\!\!\!\!\!\!
\!\!\!\!\!\!\!\!\!\!\!\!
\mathop{{\rm disc}}_{~C_{1,{\bar 1},2,\bar 2}}~\hat  E^{(0)} (\hat g) &=&
       (1- \sigma \hat g)^{1/2}\bar D(\hat g(1- \sigma \hat g)^{-3/2}),
 \label{17.cdis0}  \\
\!\!\!\!\!\!\!\!
\!\!\!\!\!\!\!\!
\!\!\!\!\!\!\!\!
\mathop{{\rm disc}}_{C_3}~\hat  E^{(0)} (\hat g)
&=&
-2i(\sigma \hat g-1)^{1/2} \nonumber \\
&&
\times\bigg[ \frac{1}{2}-
\int_{0}^\infty \frac{d \bar g'}{2\pi }
\frac{\hat g( \sigma \hat g-1)^{-3/2}}
{\bar  g'{}^2+\hat g^2(  \sigma \hat g-1)^{-3}}\bar D(-\bar g')\bigg].
\label{17.cdis}
\end{eqnarray}
For small negative $\bar g$,
the discontinuity is given by the
 semiclassical limit (see \cite{PI}, Chapter 17):
\begin{equation}
\bar D(\bar g) \approx -2i
    \sqrt{\frac{6}{ \pi }} \sqrt{\frac{4}{-3\bar g}}
      e^{4 /3 \bar g}.
\label{17.scdis}\end{equation}

We denote by
$ \varepsilon^{(0)}_k(C_i)$
the contributions of the
different cuts to the   integral
(\ref{17.moments})
for the coefficients.
After inserting (\ref{17.scdis}) into Eq.~(\ref{17.cdis0}),
we obtain from the cut along $C_1$ the semiclassical approximation
\begin{equation}
   \varepsilon ^{(0)}_k (C_1)  \approx  - 2  \, 4^k\int_{C_ 1 } \frac{d\hat
g}{2 \pi }
	\frac{1}{\hat g^{k + 1}}  \sqrt{\frac{6}{ \pi }}
	  \sqrt{-\frac{4(1- \sigma \hat g)^{5/2}}{3\hat g}}
e^{4(1- \sigma \hat g)^{3/2}/3\hat g}.
\label{17.eps}\end{equation}
For the $k$th term
 $S_k$
 of the series
 this yields an estimate
\begin{equation}
S_k\propto  \left [
\int_{C_ \gamma   } \frac{d \gamma }{2 \pi }
	 e^{f_k( \gamma )} \right]
( \sigma \hat g)^k,
\label{17.lterm}\end{equation}
where $f_k( \gamma )$ is the function of    $ \gamma \equiv \sigma \hat g$
\begin{equation}
f_k( \gamma ) = -\left( k + \frac{3}{2} \right) \log (-  \gamma )+
    \frac{4 \sigma }{3 \gamma } (1-  \gamma )^{3/2}.
\label{17.fequa}\end{equation}
 For large $k$, the integral may be evaluated via a saddle point
 approximation.
%
%
%
%
At the extremum, $\gamma  \displaystyle\mathop{\longrightarrow}_{k\rightarrow
\infty}
  \gamma _k = -{4\sigma}/{3k}$, $f_k( \gamma )$ has the value
\begin{eqnarray}
&& f_k \displaystyle\mathop{\longrightarrow}_{k\rightarrow
\infty}k\log(3k/4e\sigma)-2 \sigma.
\label{17.n21}\end{eqnarray}
The constant $-2 \sigma $ in
this limiting expression arises
when expanding the second term of Eq.~(\ref{17.fequa})
into a Taylor series,
 $(4  \sigma  /3 \gamma )(1-  \gamma )^{3/2}=
 4 \sigma /3 \gamma_k -2 \sigma  + \dots~ $.
Only the first two terms survive the large-$k$ limit.

\comment{The quadratic correction
to the saddle point produces a factor $\propto 1/k^{1/2}$.}
Thus, to leading order in $k$,
the $k$th term of the reexpanded series
			becomes
\begin{equation}
S_k\propto
e^{-2  \sigma }  \left(\frac{-3k}{e}\right)^k\left(\frac{\hat g}{4}\right)^k.
\label{17.lterm2}\end{equation}
The corresponding reexpansion coefficients
are
\begin{equation}
 \varepsilon _k^{(0)} \propto
e^{-2  \sigma } E_k^{(0)}
{}.
\label{17.prop}\end{equation}
They
have
the remarkable property of growing
			in precisely the same manner with $k$ as the
initial expansion coefficients
			$ E_k^{(0)}$, except for an overall
 suppression factor
			$e^{-2\sigma}$.
This property was discussed in Ref. \cite{PI}.

In order to estimate the convergence of the variational
perturbation expansion,
 we note that
$
\sigma \hat g =1-{1}/{ \Omega ^2}.
$
For large $ \Omega $,
this expression is
 smaller than unity.
Hence the powers $( \sigma \hat g)^k$ alone
yield
a convergent series.
An optimal reexpansion of the energy
can be achieved by choosing, for a
given large maximal order $N$ of the expansion, a parameter
$ \sigma $ proportional  to $N$:
\begin{equation}
\sigma  \approx \sigma_N\equiv cN .
  \label{17.sigman}\end{equation}
Inserting this into
(\ref{17.fequa}), we obtain for
large $k= N$
\begin{equation}
f_N( \gamma ) \approx N\left[- \log (-  \gamma )+
    \frac{4 c }{3 \gamma } (1-  \gamma )^{3/2}\right].
\label{17.fN}\end{equation}
The extremum of this function lies at
\begin{equation}
1+\frac{4c}{3 \gamma }(1- \gamma )^{1/2}(1+\frac{1}{2} \gamma )=0.
\label{17.gamma1}\end{equation}

The constant $c$ is now chosen in such a way that
the
large exponent proportional to $N$
in the exponential function $e^{f_N( \gamma )}$
due to
the first term
in (\ref{17.fN})
is canceled by an equally large contribution
from
the second term,
i.e., we require at the extremum
\begin{equation}
f_N( \gamma )=0.
\label{17.gamma2}\end{equation}
The two equations
(\ref{17.gamma1}) and
(\ref{17.gamma2})
are solved by
\begin{equation}
 \gamma
=-0.242\,964\,029\,973\,520\,\dots,~~~~c=0.186\,047\,272\,987\,975\,\dots~.
\label{17.extrema}\end{equation}
In contrast to the extremal $ \gamma $
of Eq.~(\ref{17.fequa})
which dominates the large-$k$ limit,
the extremal $ \gamma $ of the present limit, in which $k$ is also large but of
the order
of $N$,
remains finite (the previous estimate holds for $k\gg N$).
Accordingly,
the second term
$(4c/3 \gamma )(1-  \gamma )^{3/2}
$
in $f_N( \gamma )$
 contributes in full, not merely via the first
two
Taylor expansion terms of $(1-  \gamma )^{3/2}$, as it did in (\ref{17.n21}).

Since $f_{N}( \gamma )$ vanishes at the extremum,
the $N$th term in the
reexpansion
has the order of magnitude
\begin{equation}
 S_N\propto
 ( \sigma_N \hat g_N )^{N}
=  \left (1-\frac{1}{ \Omega _N^2}\right  )^{N}.
\label{17.lastterm}\end{equation}
According to
(\ref{17.sigman}), the frequency
$ \Omega _N$ grows for large $N$ like
\begin{equation}
 \Omega _N \sim
\sigma _N^{1/3}g^{1/3} \sim
(cNg)^{1/3}.
\label{17.omegan}\end{equation}
As a consequence, the last term
of the series decreases for large $N$ like
\begin{equation}
  S_N(C_1)\propto
  \left[1 -
      \frac{1}{ ( \sigma _N g)^{2/3}}
\right]^N
\approx e^{ - N/( \sigma g) ^{2/3}}
\approx e^{ - N^{1/3}/(cg)^{2/3} }.
\label{17.sn}\end{equation}

This estimate does not yet explain the convergence of the
variational perturbation expansion in the strong-coupling limit observed in
 Figs. \ref{slope1} and \ref{janke12}.
For the contribution of the cut $C_1$ to $S_N$,
the derivation of such a behavior
requires including
a little more information
into the estimate.
This information is supplied by
 the empirically observed property, that the
best $ \Omega _N$-values
lie
for finite $N$ on a curve (see Chapter 5 in \cite{PI}):
\begin{equation}
 \sigma _N \sim  cN\left (1 + \frac{6.85}{N^{2/3}}\right).
\label{17.omeganc}
\end{equation}
Thus the asymptotic behavior
		(\ref{17.sigman})
		receives,
at
a finite $N$, a rather large correction.
By inserting this $ \sigma _N$ into  $f_N( \gamma )$ of (\ref{17.fN}),
we find an extra exponential factor
\begin{eqnarray}
e^{ \Delta f_N}&\approx&
\exp\left[N\frac{4c}{3}\frac{(1- \gamma )^{3/2}}{ \gamma
}\frac{6.85}{N^{2/3}}\right]\nonumber \\
&=&\exp\left[-N\log(- \gamma )\frac{6.85}{N^{2/3}} \right] \approx
e^{-9.7N^{1/3}}.
\label{x17.389}\end{eqnarray}
This reduces the size of the last term
due to the cut $S_N(C_1)$
in
(\ref{17.sn}) to
\begin{equation}
  S_N(C_1)\propto
e^{ -[ 9.7+(cg)^{-2/3}]
N^{1/3}
},
\label{17.sn2}\end{equation}
which agrees with the convergence seen in
 Figs. \ref{slope1} and \ref{janke12}.

Note that there is no need to evaluate the effect of
the shift in the extremal value of $ \gamma $
caused by the correction term in (\ref{17.omeganc}), since
this would be of second order
in $1/N^{2/3}$.

\begin{figure}[tbhp]
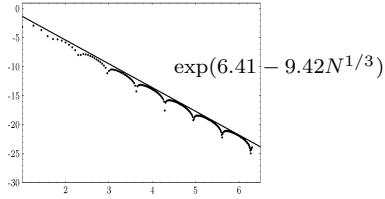

\input  poscappl
\caption[Theoretically obtained convergence behavior of the $N$th approximants
$S_N$ for $ \alpha _0$
]{Theoretically obtained convergence behavior of the $N$th approximants
$S_N$ for $ \alpha _0$,
to be compared with the empirically found
 behavior in Fig. 1.
}
\label{poscappl}\end{figure}

How about the contributions of the other cuts?
For $C_{\bar 1} $, the integrals in
(\ref{17.moments}) runs from $\hat g=-2/ \sigma $ to $-\infty$
and decrease like $(-2/ \sigma )^{-k}$. The associated last term
$S_N(C_{\bar 1})$ is of the negligible order $e^{-N\log N}$.
For the cuts $C_{2,\bar 2,3}$, the
integrals
(\ref{17.moments}) start at $\hat g=1/ \sigma $ and have therefore
the leading behavior
\begin{equation}
\varepsilon _k^{(0)}(C_{2,\bar 2,3})\sim  \sigma ^{k}.
\label{}\end{equation}
This implies a contribution to the $N$th term in the reexpansion
of the order of
\begin{equation}
S_N(C_{2,\bar 2,3})\sim ( \sigma \hat g)^N,
\label{}\end{equation}
which decreases merely like
(\ref{17.sn}) and does not explain the empirically observed
convergence in the strong-coupling limit.
As before, an additional information
produces a better estimate.
The cuts in Fig.~\ref{17.contour}
do not really reach the point
$ \sigma \hat g=1$.
There exists a small circle of radius $ \Delta \hat g>0$
in which $\hat E^{(0)}(\hat g)$ has no singularities at all.
This is a consequence of the fact
unused up to this point that
the strong-coupling expansion
(\ref{5.scexpag})
converges for
$
g>g_{\rm s}.
$
For the reduced
energy,
this expansion
reads:
\begin{eqnarray}
&&\!\!\!\!\!\!\!\!\!\!\hat E^{(0)}(\hat g)=  \left(
\frac{\hat g}{4}\right)^{1/3}\left\{ \alpha _0
+ \alpha _1\left[\frac{\hat g}{4}\frac{1}{(1-  \sigma \hat
g)^{3/2}}\right]^{-2/3}
\!\!\!\!\!\!+ \alpha _2\left[\frac{\hat g}{4 }\frac{1}{(1-  \sigma \hat
g)^{3/2}}\right]^{-4/3}
\!\!\!\!\!\!+\dots\right\}
\!.                    \nonumber \\&&~~
\label{5.scexpaghat}\end{eqnarray}
The convergence of
(\ref{5.scexpag}) for $g>g_{\rm s}$
implies that (\ref{5.scexpaghat})
converges for all
$  \sigma  \hat g$ in a neighborhood of the point
$  \sigma  \hat g=1$ with a radius
\begin{equation}
  \Delta (\sigma  \hat g)\sim \left(\frac{\hat g}
{-\bar g_{\rm s}}\right)^{2/3}=
\left\{\frac{1}{- \sigma \bar g_{\rm s}}[1+ \Delta (\sigma  \hat
g)]\right\}^{2/3}
\label{17.zusatz}\end{equation}
where $\bar g_{\rm s}\equiv g_{\rm s}/ \omega ^3$.
For large $N$,$ \Delta (\sigma  \hat g)$ goes to zero like $1/(N|\bar gs|
c)^{2/3}$.
Thus the integration contours of the moment integrals
(\ref{17.moments})
for the contributions
 $ \varepsilon ^{(0)}_k (C_i)$ of the other cuts
do not begin at the point
$ \sigma \hat g=1 $, but
a little distance
$ \Delta  (\sigma   \hat g)$
away from it.
This generates an additional
suppression factor
\begin{equation}
 \left({ \sigma  \hat g}\right)^{-N}\sim \left[1+\Delta (\sigma  \hat g)
\right]^{-N} .
\label{}\end{equation}
Let us set $-\bar g_{\rm s}=|\bar g_{\rm s}|\exp(i \varphi_{\rm s})$ and
$x_{\rm s}\equiv (-\hat g/\bar g_{\rm s})^{2/3}=-| x_{\rm s} |\exp(i \theta )$,
and introduce the parameter $a\equiv 1/ [|\bar g_{\rm s}| c]^{2/3}$.
Since there are two complex conjugate contributions
we obtain, for large $N$ a last term of the order of
\begin{equation}
S_N(C_{2,\bar 2,3})\approx
e^{-N^{1/3}a\cos \theta}\cos(N^{1/3}a\sin \theta).
\label{}\end{equation}
By choosing
\begin{equation}
|\bar g_{\rm s}|\sim 0.160,~~~~\theta\sim -0.467,
\label{17.gval}\end{equation}
we obtain the curves shown in Figs. \ref{poscappl} and \ref{apprscc3},
which agree
very well with the observed
 Figs. \ref{slope1} and \ref{janke12}. Their envelope
has the asymptotic falloff
$e^{-9.23N^{1/3}}$.

\begin{figure}[tbhp]
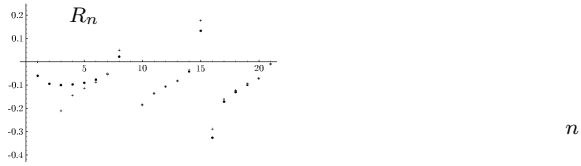

\input  strccomp
\caption[Comparison of ratios between successive expansion coefficients
$R_n$
of the strong-coupling expansion and of the ratios
$R_n^{\rm as}$
of the expansion of
a superposition of two
square root singularities at $g=0.156\times\exp(\pm 0.69)$]{
Comparison of ratios $R_n$ between successive expansion coefficients
of the strong-coupling expansion (dots) with the ratios $R_n^{\rm as}$
 of the expansion of
a superposition of two
singularities at $g=0.156\times\exp(\pm 0.69)$ (crosses).
}
\label{strccomp}\end{figure}

Let us see how the positions of the largest
Bender-Wu singularities compare with what we can extract directly from
the strong-coupling series (\ref{5.scexpag}) up to order 22 \cite{jk}.
For a pair of square root singularities at
$x_{\rm s}=-|x_{\rm s}|\exp(\pm i \theta ) $,
the coefficients of a power series
$\sum  \alpha _n x^n$  have the asymptotic ratios
$R_n\equiv  \alpha _{n+1}/ \alpha _n\sim
 R_n^{\rm as} \equiv
-\cos[(n+1)  \theta  + \delta ]/
|x_{\rm s}|
\cos(n  \theta  + \delta ) $.
Plotting the ratios $R_n$
for the coefficients $ \alpha _n$ in Fig. \ref{strccomp} we see that
for large $n$, they are well reproduced by
$ R_n^{\rm as}$
if we choose
\begin{equation}
|x_{\rm s}|=1/0.117,~~~ \theta =-0.467,
\label{}\end{equation}
with an  irrelevant phase angle
$ \delta =-0.15$. The angle $ \theta $ is in excellent agreement with
the value
found in (\ref{17.gval}).
{}From $|x_{\rm s}|$
we find
 $|\bar g_{\rm s}|=4|1/x_{\rm s}|^{3/2}=0.160$, again
in excellent agreement with
(\ref{17.gval}).

\begin{figure}[tbhp]
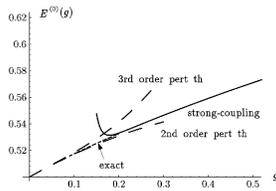

\input  strconv
\caption[Strong-Coupling Expansion for the ground-state energy
in comparison with the
exact values and the perturbative results of 2nd and 3rd order]{
Strong-Coupling Expansion for the ground-state energy
in comparison with the
exact values and the perturbative results of 2nd and 3rd order.
The convergence radius  in $1/g$ is larger than $1/0.2$.
}
\label{strconv}\end{figure}

Note that this convergence radius is compatible with the heuristic convergence
of the
strong-coupling series up to order 22, as can be seen
in Fig. \ref{strconv} by comparing
the series with the exact curve.

While this work was in progress, we received a Genova preprint by
R. Guida, K. Konishi, and H. Suzuki (hep-th/9505084)
in which the exponentially fast
convergence found in Ref. \cite{jk}
is proved rigorously by reducing it to the known
convergence
of the strong-coupling expansion.
\\~~\\
%


%
\newpage
\begin{figure}[tb]
\input contour
\caption[Cuts in the complex $\hat g$-plane
whose moments
with respect to the inverse coupling constant
determine the reexpansion coefficients]{Cuts
in the complex $\hat g$-plane
whose moments
with respect to the inverse coupling constant
determine the reexpansion coefficients. The cuts inside the
shaded circle happen to be absent
due to the convergence of the strong-coupling expansion
for $g>g_{\rm s}$ (from Ref. \cite{PI}.}
\label{17.contour}\end{figure}
\begin{figure}[tbhp]
\input  apprscc3
\caption[Theoretically obtained oscillatory behavior around the asymptotic
approach
of $ \alpha _0$ to its exact value
as a function of the order $N$ of the approximant]{
Theoretically obtained oscillatory behavior around the asymptotic approach
of $ \alpha _0$ to its exact value
as a function of the order $N$ of the approximant,
 to be compared with the empirically found
 behavior in Fig. 2, averaged between even and odd orders.
}
\label{apprscc3}\end{figure}
\vspace{-1cm}
\end{document}